\documentclass[%
reprint,
 amsmath,amssymb,
aps,
pra,
]{revtex4-1}

\usepackage{graphicx}
\usepackage{dcolumn}
\usepackage{bm}

\begin{document}


\title{Measurement of the scalar polarizability within the   5P$_{1/2}\rightarrow \textbf{6S}_{1/2}$ \\  410 nm transition in atomic indium}
\author{G. Ranjit, N. A. Schine, A. T. Lorenzo\footnote{current address: The University of Arizona, College of Optical Sciences, 1630 E. University Blvd., Tucson, AZ 85721 }, A. E. Schneider, and P. K. Majumder}
\email{pmajumde@williams.edu}
\affiliation{Department of Physics, Williams College, Williamstown, MA 01267}

\date{\today}

\begin{abstract}
We have completed a new measurement of the Stark shift in $^{115}$In within the 410 nm 5P$_{1/2} \rightarrow 6S_{1/2}$ transition.  We measure the Stark shift constant to be $k_S= - 122.92(33)$ kHz/(kV/cm)$^{2}$, corresponding to a difference in the 6S$_{1/2}$ and 5P$_{1/2}$  state polarizabilities, $\Delta\alpha_0$, of $1000.2 \pm 2.7 \,a_0^3$ (in atomic units).  This result is a factor of 30 more precise than previous measurements and is in excellent agreement with a new theoretical value based on an \emph{ab initio} calculation of the wave functions in this three-valence-electron system.  The measurement was performed in an indium atomic beam apparatus, used a GaN laser diode system, and exploited an FM spectroscopy technique to extract laser transmission spectra under conditions where our interaction region optical depth was typically less than 10$^{-3}$.  
\end{abstract}

\pacs{32.60.+i, 32.70.Jz, 32.30.Jc, 32.70.Cs}
\maketitle


\section{\label{sec:level1} Introduction}

High precision atomic structure measurements have long been an important tool in testing the accuracy and guiding the refinement of theoretical techniques aimed at calculating atomic wave functions in multi-electron atomic systems.   Alkali systems have played a particularly important role in this experiment-theory interplay.  More recently, new \emph{ab initio} calculational techniques have produced theoretical results of improved precision for multi-valence-electron systems\cite{Saf09, Saf07, Saf06, SahooDasLifetime}.  Of relevance here are the Group III systems such as indium and thallium.  Thallium, in particular, has played an important role in tests of discrete symmetry violation over recent decades\cite{Regan,Vetter,PorsevSafEDM-Tl}, and a proposal to measure the permanent electric-dipole moment of indium was recently published\cite{SahooEDM-In}.  The size of these symmetry-violating observables scales rapidly with the atomic number, encouraging the use of high-Z systems.  This therefore requires independent, precise atomic wave function calculations in order to distinguish quantum mechanical effects from the elementary particle physics observables being targeted.  For example, the present  2-3\% uncertainties in \emph{ab initio} wave function calculations in thallium currently limit the quality of the standard model test provided by a 1995 thallium parity nonconservation measurement\cite{Vetter}.  A very similar theoretical method to that used for thallium can be applied to other three-valence systems such as indium and gallium\cite{Saf07}. 

Recently our group  measured the hyperfine constants of the indium $6P_{3/2}$ excited state\cite{Gunawardena09}, providing a test of short-range electron wave function behavior as well as a measurement of the nuclear quadrupole moment.   The Stark shift result reported here is quite complementary to this hyperfine structure work in that it provides a test of long-range wave function properties.  Previous experimental work to measure indium polarizability includes a 1970 atomic beam spectroscopy experiment\cite{FowYel}, and a 1984 ground state measurement based on atomic beam deflection\cite{Guella84}, both at the 10\% level of accuracy.  

Our new polarizability measurement, with its 0.3\% accuracy provides a  benchmark test of two distinct, high-precision calculational methods which can be applied to multi-valence atoms such as indium.  One method, employing a coupled-cluster approach, has been particularly effective for mono-valent systems\cite{SafJohn08}, but also produced results\cite{Saf06} in excellent agreement with experimental polarizability measurements in thallium\cite{Doret02,Demille94}.  A second theoretical strategy incorporates a configuration interaction approach to better accommodate three-particle states.  Both of these methods were very recently employed to compute the $6S_{1/2}$ and $5P_{1/2}$ state polarizabilities in indium.  The agreement between these theoretical methods and between theory and our experimental result is very good, and is detailed in \cite{Saf13}.  Finally, as discussed at the end of this manuscript, our experimental result and the theoretical expressions can be combined to extract new, accurate values for the indium $6P_{1/2}$ and $6P_{3/2}$-state lifetimes.


\section{Atomic Structure Details}
We focus on the principal naturally-occuring isotope of indium ($^{115}$In, 96\% abundant), which in our Doppler-narrowed atomic beam geometry is spectroscopically isolated from the small $^{113}$In component. $^{115}$In has nuclear spin $I=9/2$, so that both the $5P_{1/2}$ and $6S_{1/2}$ states 
contain $F=5$ and $F=4$ hyperfine levels.  The respective 11.4 and 8.4 GHz hyperfine 
splittings (HFS) of the ground and excited states again
yield an entirely resolved spectrum in our 
atomic beam apparatus.  Because we study a $J=1/2\rightarrow J=1/2$ 
transition, the Stark shift of each level has only a scalar 
component, producing a common shift of all sublevels within a 
given state, and yielding an experimental result which is independent 
of relative laser and static electric field polarization.  
Expressing the energy shift of a given level as 
$\Delta W = - \frac{1}{2} \alpha_{0} E^{2}$, where $\alpha_{0}$ is the 
scalar polarizability, the observed frequency shift of the 
410  nm line can then be expressed as: $\Delta\nu_{S} = - \frac{1}{2h} [\alpha_{0}(6S_{1/2}) - 
\alpha_{0}(5P_{1/2})] E^{2}$.  In the one electron 
central field (OECF) valence approximation, the polarizability is calculable from second-order perturbation theory, and for the state '$v$' is given by the expression\cite{Hunter88,MitroySaf10}
\begin{eqnarray}
\alpha_0(v) & = & \frac{2}{3(2j_{v}+1)} \sum\limits_{k}  \frac{\langle k \| D \| v \rangle^2}{E_k - E_{v}},
\label{alphasum}
\end{eqnarray}
where $\langle k \| D \| v \rangle$ is the reduced electric-dipole (E1) matrix element, and the sum extends over all states for which the E1 matrix elements are non-zero.

Due to the size of the HFS  compared to the optical 
transition frequencies ($\delta\nu_{HFS} / 
\delta\nu_{opt}$  of order 
$10^{-5}$) we expect no measurable difference in 
Stark shift value among four
hyperfine component lines, given our level of 
experimental precision.  Small tensor components to the Stark shift of 
individual sublevels induced by
higher-order hyperfine interaction effects are similarly of 
negligible magnitude in our experiment\cite{Gould76}.  As discussed below, we did acquire data for two different hyperfine transitions, and different laser linear polarization directions, as part of our overall search for potential systematic errors. 

In this experiment, we perform transverse laser spectroscopy.  Given the geometry of our atomic beam and interaction region, we find that the residual Doppler broadening of our atomic spectrum is reduced roughly 15-fold from the value that would be found in our heated source.  As such, the absorption line shape is described by a Voigt profile with a 23 MHz Lorentzian component due to the natural line width, and a roughly 100 MHz Gaussian component due to the residual Doppler broadening.  As described below, we use an electro-optic modulator to add FM sidebands to the spectrum of our blue laser.  We  demodulate the subsequent atomic beam transmission signal at either the first or second harmonic of the modulation frequency. The resultant atomic line shapes can be computed as a function of FM modulation index, choice of demodulation frequency, and relative phase\cite{bjorkFM, james94}.   As discussed in Sec. IV, we have developed a model for exact fitting of our RF-demodulated spectra, but also perform simplified polynomial or Lorentzian fits to the central portion of our line shapes to extract line centers and Stark shift information.

\section{Experimental Details}
\subsection{Atomic beam system}
Fig. 1 shows a cross section of our atomic beam oven source, vacuum chamber, and interaction region.  Our atomic beam source consists of a cylindrical molybdenum crucible capable of containing 100 grams of metallic indium. The crucible is supported by two legs that tilt the cylinder at roughly $20^o$ upwards from the horizontal. The open front face of the cylinder is milled so that it is in the vertical plane, and we cap the crucible with a molybdenum faceplate containing 20 parallel  0.25-mm-wide vertical slits extending over a 2 cm horizontal width. The 0.5 cm thickness of the faceplate serves to help pre-collimate our atomic beam by creating a series of parallel 'tunnels' from which the indium atoms must effuse. The crucible sits within an alumina tube around which we attach a pair of resistive heaters. These are surrounded by layers of carbon felt insulation and thin stainless steel (serving as a heat shield). A water-cooled cylindrical copper cold wall forms the outermost layer of our source oven. The whole assembly rests on an adjustable base that bolts to the floor of the vacuum chamber. Four thermocouples attached to crucible, heater, stainless steel heat-shield and copper cold-wall allow for continuous temperature monitoring.  To reach our target temperatures between 900 and 1000$^o$C, we require roughly 0.6 kW of (AC) power.  We regulate the temperature of the crucible using a variable-duty-cycle controller which can produce stable source temperatures up to 1100$^o$C.  
\begin{figure}
\includegraphics[scale = 0.160]{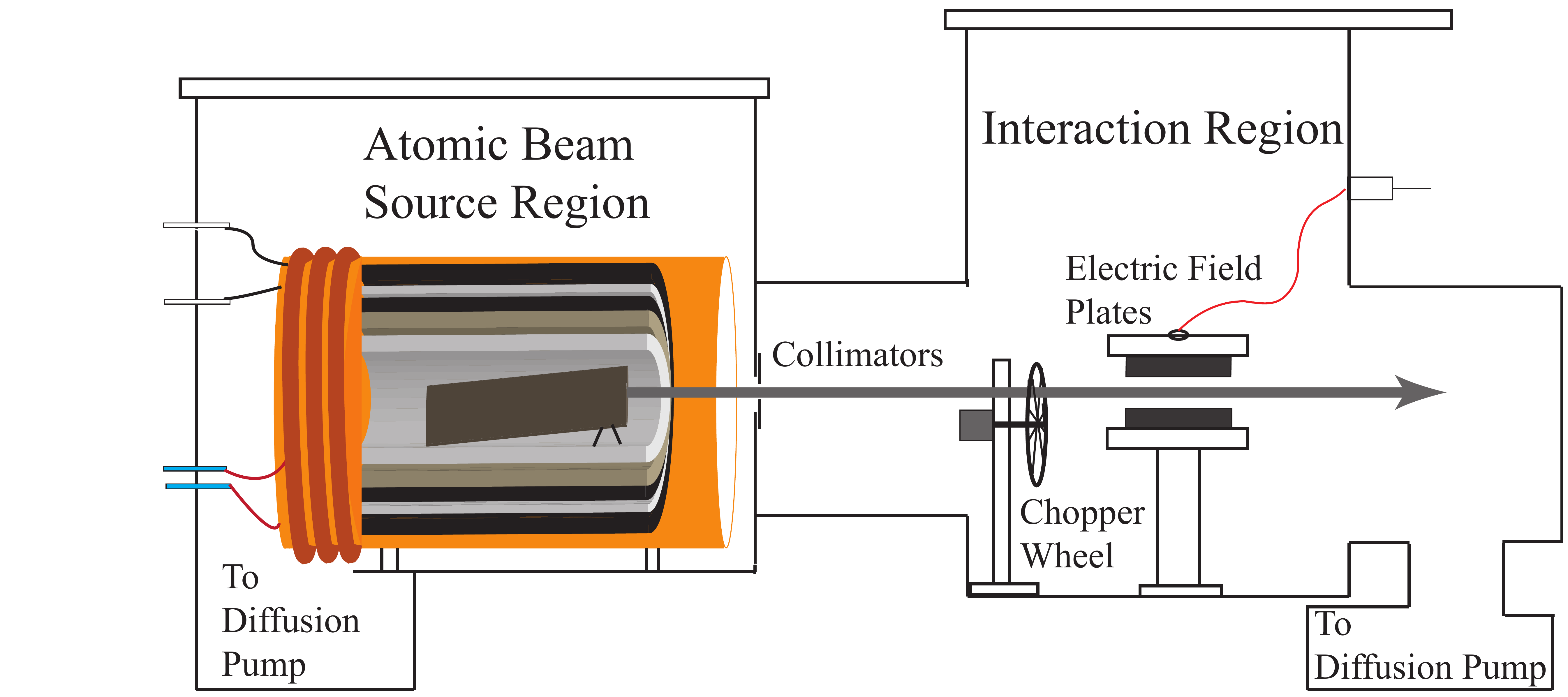}
\caption{A cross-sectional view of the atomic beam source, vacuum chamber, and interaction region.  The source-interaction distance is roughly 40 cm.}
\end{figure}

Our stainless steel vacuum chamber includes a pair of liquid-N$_2$-trapped diffusion pumps (Varian M-6), one located beneath the source oven, and one beneath the interaction region chamber. Under hot-source-oven operating conditions, we typically observe 1x10$^{-6}$ torr pressure in our source chamber, while the pressure in the differentially-pumped interaction region is 3 or 4 times lower.  Roughly half way along the  $\sim$ 40 cm source-interaction region path we place adjustable razor blade collimators to define and collimate our atomic beam in both vertical and horizontal directions, creating a 2-mm-high, 1.5-cm-wide ribbon-shaped atomic beam.  Just upstream of our electric field plates we insert one additional collimation iris of comparable dimensions to minimize the amount of scattered metallic indium which might coat the field plate surfaces.  Finally, we attach an in-vacuum chopping wheel to the final collimator assembly, just upstream of our interaction region.  In this way, we can modulate the atomic beam at a typical frequency of 500 Hz, leading to removal of background features in our atomic spectra as discussed below.  A set of three pairs of orthogonal magnetic-field-cancellation coils on the outside of the interaction chamber is used to reduce the ambient magnetic field to $<10^{-6}$ T within the interaction region.
		
\subsection{Electric field plates and high voltage system}

Our electric field plate assembly consists of a pair of highly polished circular stainless steel plates that are 10 cm in diameter. The two plates are separated by four ceramic spacers and positioned horizontally on a ceramic pedestal. The separation between the plates was repeatedly measured using metric gauge blocks to be 0.999(1) cm.  The modest temperature increase characteristic of the 
downstream interaction region during beam operation causes negligible change in plate separation, 
especially given that expansion of the ceramic posts and steel plates  tend to oppose one another.
Our blue laser beam intersects the 1.5-cm-wide 
atomic beam at the center of the plate assembly.  Independently, we 
have performed simulations to insure that, given our interaction 
region geometry, any non-idealities due to 
the finite size of the plates would affect our electric field 
calibration only at the level of 1 part in 10$^{5}$ or below. 

A high voltage power supply (Glassman  ER40P07.5) allows 
the application of up to 40 kV to the plates.  Our data acquisition 
system allows computer control of both polarity and magnitude of this 
voltage.  In series with the HV output and the field plates, we place 
both a chain of ballast resistors of total resistance 100 M$\Omega$ (for 
current limiting and power supply protection), as well as a 1 
M$\Omega$ resistor between the ground plate and the power supply 
return (for leakage current monitoring).  Leakage currents were typically in the nA range,  so that errors due to leakage current-induced 
voltage drops were negligible for our purposes.   In parallel with the power supply we installed a 
commercial high voltage divider (Ross Engineering, Inc.).  
This device has a quoted precision of 0.01\%, a total resistance of 180 
M$\Omega$, and is customized to be impedance matched to the particular voltmeter we use 
(Keithley model 197A, 0.011\% absolute voltage measurement accuracy).  The data acquisition program 
reads the output of the Keithley voltmeter via a GPIB interface, and, to the 
quoted precision of the instruments, this is exactly 1/1000 of the 
actual high voltage applied. 

\subsection{Optical System}
The experiment uses a commercial external cavity GaN diode laser system in the Littrow configuration (Toptica Photonics, DL 100) to produce roughly 10 mW of light at a wavelength near 410.2 nm. The laser can be tuned to excite any of the four hyperfine transitions in indium 5P$_{1/2} $(F = 4, 5) $\rightarrow$ 6S$_{1/2}$ (F$'$ = 4, 5) manifold. After passing through an optical isolator, a fraction of the laser output is directed into a Doppler-free saturated absorption system.  This system includes an optical chopping wheel which interrupts the pump beam, and we record the subsequent lock-in amplifier output of the probe beam signal.  We do not lock our laser using this signal, but instead use the Doppler-free absorption peak as a frequency reference to facilitate comparison of the spectral shift in atomic beam transmission signal when we turn on and off the electric field in our interaction region. \\

The main laser beam is then passed through an electro-optic modulator (EOM; New Focus, Model 4001). The EOM is driven by a 100 MHz RF synthesizer and power amplifier. We obtain a frequency modulated (FM) laser spectrum whose exact sideband distribution is determined by the EOM input power.   A portion of the modulated laser beam is then directed to a 1 GHz free-spectral-range confocal Fabry-Perot (FP) Interferometer  (Burleigh RC-110) whose finesse is roughly 40.  We isolate and passively temperature stabilize the FP cavity by covering it with acoustical insulation material. By detecting the FP transmission signal as we scan our laser frequency, the FM laser spectrum, with its precisely-known 100 MHz sidebands, can thus be monitored, and the laser frequency scan can be calibrated and linearized.  In order to create a spectrum with measurable second and third-order FM sidebands, we drive the EOM with sufficient power to achieve modulation depths, $\beta$, ranging from approximately 1.0 to 3.0. 

The modulated laser beam is then directed into and through our atomic beam apparatus via windows which are anti-reflection coated for our 410 nm wavelength. As it enters the vacuum chamber, the laser beam has a 1-mm diameter,  an adjustable linear polarization, and is arranged to intersect the atomic beam in a transverse fashion.  By passing our laser beam through 1-mm-diameter collimating apertures on either side of our vacuum chamber, we insure that the we have a reproducible, well-defined interaction geometry.  This minimizes any possible Doppler-shift-related systematic  errors in our spectra. After exiting the vacuum chamber, having interacted with the atomic beam, we collect the optical transmission signal on a 1-GHz-bandwidth photodiode (New Focus, model 1601). Fig. 2 shows a complete schematic layout of our optical and signal processing arrangement.
\begin{figure}[t]
\includegraphics[scale=0.9]{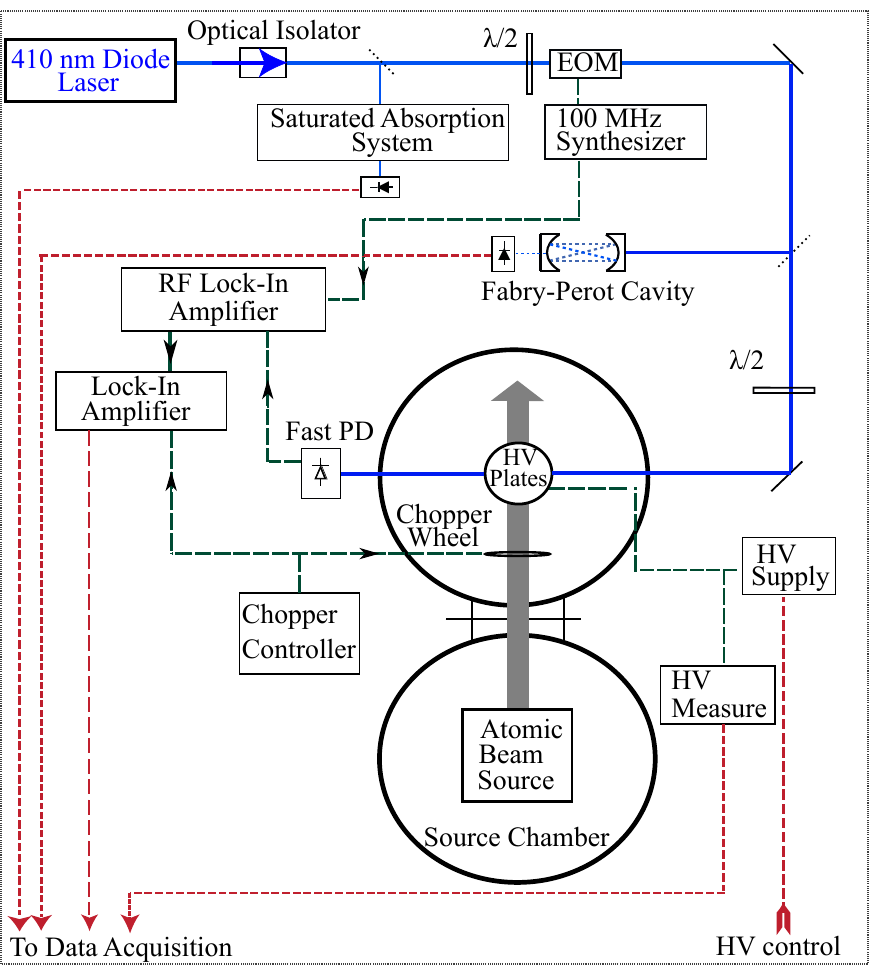}
\caption{\label{fig:epsart} A complete  experimental setup indicating the configuration of the Doppler-free saturated absorption system (top),  Fabry-Perot interferometer, and transverse atomic beam FM spectroscopy system (below).}
\end{figure}

\subsection{FM signal processing and lock-in detection}

By applying a linear voltage ramp to the piezoelectric transducer (PZT) of the diode laser external cavity, we scan the 410 nm laser frequency upwards and downwards over  $\sim 1.5$ GHz which includes either the  F = 5 $ \rightarrow $ F$'$ = 4 or   F = 4 $ \rightarrow $ F$'$ = 5 hyperfine resonance.   Because the vapor pressure of indium is comparatively low (for example, 1000 times lower than thallium in this temperature range), we observe an atomic beam optical depth of approximately 0.001 in our interaction region even at oven temperatures near 1000$^o$C.   We have therefore made use of an FM spectroscopy / RF detection technique which greatly improves the quality of our atomic absorption spectra under conditions where direct transmission measurement is not a feasible option.  As described below, we demodulate the transmitted atomic absorption signal at either the fundamental FM frequency (`1f' = 100 MHz in our case) or at the second-harmonic frequency (2f).  One can also study the in-phase or the quadrature component of the demodulated signal. We find that higher FM modulation depth complicates the line shape, but provides us with a larger signal size.  A complete analysis of the form of the expected demodulated signals under various conditions can be found elsewhere\cite{bjorkFM,james94}.  
The frequency-modulated electric field can in general be expressed in terms of sideband components with relative amplitudes given by Bessel functions, $J_n$, whose arguments are the modulation depth. Defining $\omega_c$ and $\omega_m$ to be the carrier and modulation frequency respectively, and $\beta$ to be the modulation depth, we have:
\begin{eqnarray}
E(t) && = E_0 \exp[i (\omega_c t + \beta \sin \omega_m t)]  \nonumber\\
 && =  E_0 \exp(i \omega_c t) \sum\limits_{n=-\infty}^{+\infty} J_n(\beta) \exp(i \omega_m t)
 \end{eqnarray}
 
 After interacting with the atomic beam, we can associate a (complex) electric field transmission function, $T_n(\omega_c+n\omega_m)$, with each sideband component.  In the limit of very small absorption, we approximate this as: $T_n \approx 1 - (\delta_n + i \phi_n)$, where $\delta$ is the frequency-dependent atomic absorption line shape and $\phi$ is the associated (dispersive) phase shift.  The transmission signal we detect is then proportional to the absolute square of the                                                                                                                                                                                                                                                                                                                                                                                                                                                                                                                                                                                                                                                                                                                                                                                                                                 transmitted field.  Our RF lock-in amplifier (SRS model SR844) is referenced to the synthesizer which drives the EOM.  By detecting the Fourier component of the transmitted intensity at either 1f or 2f, we pick out particular cross terms in the square of the sum above.  We can derive analytic expressions for the in-phase and quadrature component demodulated spectra of both the 1f and 2f demodulated signals for arbitrary modulation depth.  For example, for the case of demodulation at the fundamental modulation frequency, we find\cite{james94}:
 \begin{eqnarray}
 I_{[1f]}(\omega_c)  \, &&  \propto \sum\limits_{n = 0}^{\infty}J_nJ_{n+1} \ [(\delta_{-n-1} -\delta_{n+1}+\delta_{-n}-\delta_{n}) \cos\theta_d  \nonumber\\
 && + (\phi_{-n-1} -\phi_{-n}+\phi_{n+1} -\phi_{n})\sin\theta_d \ ] 
\label{eq1f}
\end{eqnarray}
 where $\theta_d$ is the relative phase between the FM signal and the lock-in detector. Each absorption component $\delta_n(\omega_c+n\omega_m)$ represents a distinct, frequency-shifted Voigt profile.  We use an appropriately truncated sum of this form to fit our demodulated spectra, where the modulation depth and lock-in phase are left as fit parameters in our analysis.
 
Eq. \eqref{eq1f} predicts a zero-background demodulated signal.  Because the absorption in our atomic beam is so small, in practice we find that our demodulated spectra contain some non-zero frequency-dependent background in addition to our atomic absorption signal, likely due to some purely optical effects, such as etaloning due to small reflections within the EOM itself.  We can completely eliminate this residual optical background pattern using our atomic beam chopping wheel.  We direct the output signal of our RF lock-in amplifier to a second low-frequency lock-in amplifier (SRS model SR810) which is referenced to the chopping wheel rotation frequency.  The output of this second-stage demodulation is a truly background-free line shape which we record and analyze.
 

\subsection{Data acquisition and experiment control}
\begin{figure}
 \includegraphics[scale = 0.8]{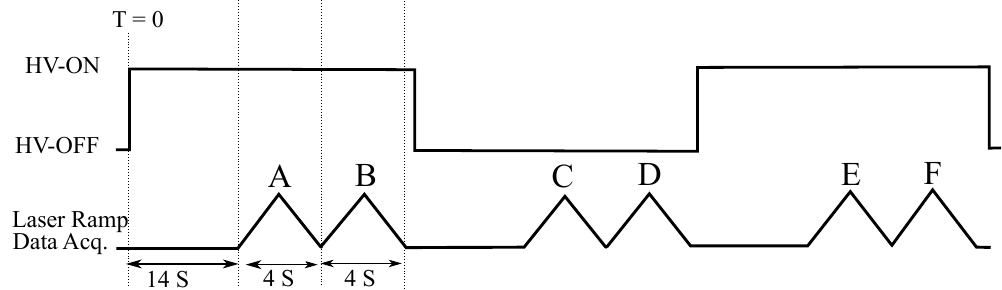}
\caption{A timing schematic for the experiment and data acquisition. We compute a set of Stark shift values by comparing HV on-off  scans, such as B-C above, and a separate set of values by comparing HV off-on scans such as D-E in the figure.}
\label{timingshceme} 
\end{figure}

The high voltage (HV) is programmed and controlled using a LabVIEW program. A computer-generated timing sequence initiates switching on or off the high voltage. After each state switch, we wait roughly 14 seconds for all high-voltage-switching transients to subside. This delay time was chosen by studying the consistency of scans acquired with various shorter delay times.  We then initiate a triangular external voltage ramp signal which we apply to the PZT in the laser's external cavity.  Two upward and downward going sweeps, each requiring roughly 3 seconds, are completed for each high voltage state.  In this way, we can compute the Stark shift by comparing consecutive laser scans in two different ways.  By comparing the [HV on - HV off] frequency shift to that obtained with the temporal order reversed we can study and potentially eliminate any slow drift, or HV state-specific effect in the experimental system that might lead to a systematic error in our measured frequency shift. In all cases, we record and analyze upward- and downward-going laser scans separately.  Fig. \ref{timingshceme} summarizes the data acquisition timing sequence.  The  FM signal, FP transmission signal, Doppler free saturated absorption (SA) signal, and the output of the meter which monitors the field plate high voltage are sent to our data acquisition interface and recorded by the LabVIEW program.

\section{Data and Analysis}
\subsection{Summary of data collected}
We typically acquired alternating HV off / HV on scans in the sequence described above using a particular choice of HV for a 30-minute period.  The actual voltage measured by our high-precision Keithley voltmeter is recorded for each data point, though in practice the voltage value varies only in the third decimal place from cycle to cycle within a set of nominally identical HV runs.  This produced $\sim$ 200 individual spectra, providing $\sim$ 100 Stark shift determinations, of which half are upward-going laser scans, and half downward scans. We then changed our nominal value for the HV, repeating these half hour data collection cycles for 11 values of HV between 10 kV and 20 kV.  We analyzed 18 such days of data in total.  A wide variety of experimental conditions were changed over the course of the full data-collection period, including hyperfine transition, laser polarization and power, EOM power (modulation depth for FM spectrum), and RF lock-in detection frequency (1f, 2f). In total, we obtained roughly 6,000 individual Stark shift measurements which were analyzed in multiple ways to insure that the Stark shift values extracted from the fits were consistent and reliable.  The final statistical error in our measurement is less than 0.1\%.  The high statistical precision thus allowed us to explore many possible systematic error contributions to the data set, as we discuss  below.

\subsection{Data fitting procedure}
\begin{figure}
 \includegraphics[scale = 0.25]{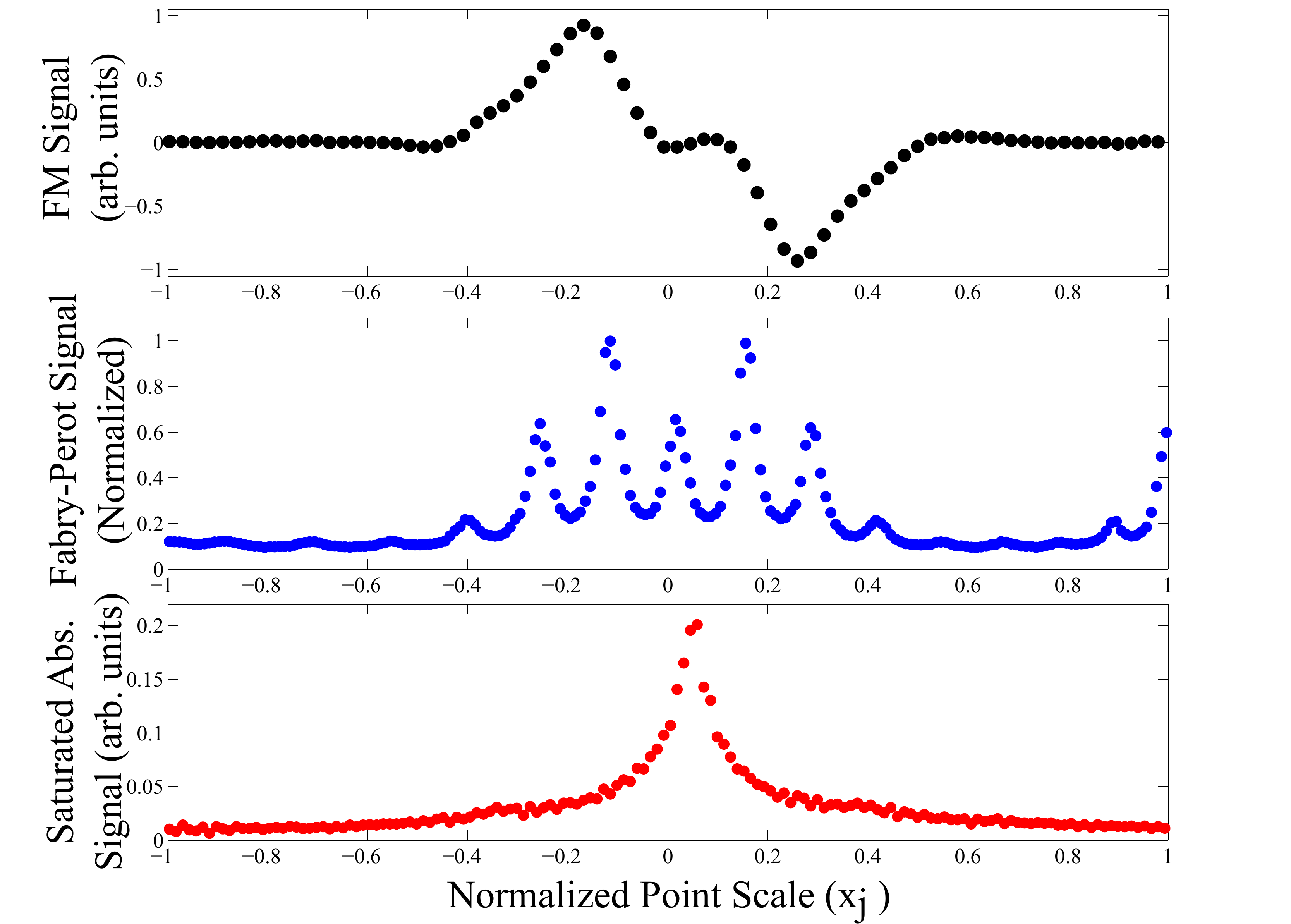}
\caption{\label{fig:epsart} Simultaneous data acquisition of demodulated FM signal (top), Fabry-Perot transmission, and saturated absorption signal (bottom).  For this scan, we demodulate the FM signal at 1f, and the entire scan extends over roughly 1.5 GHz.}
\label{simultaneous}
\end{figure}

Since in this experiment we compare atomic resonance positions after switching HV state, we require that the system remain stable over time scales of at least tens of seconds.  Furthermore, we require a spectral feature that can be used as a frequency reference point.  Finally, as we ramp the voltage applied to the laser cavity PZT for our scans, we must linearize and calibrate our scan to produce an accurate frequency axis.  For each laser scan, we simultaneously collect the demodulated FM atomic beam signal, the FP transmission signal, and the saturated absorption signal from our supplementary vapor cell.  Fig. \ref{simultaneous} shows such a single scan and the associated signals that we collect.  For this data scan we employed 1f demodulation with an FM modulation depth of roughly 1.5.  We typically scan the laser over a $\sim$ 1.5 GHz range centered on the atomic resonance.  To begin a data collection cycle, we manually tune the FP cavity length to insure that the FM spectrum of the FP cavity is centered on the atomic resonance, and we find that we do not need to retune the cavity for several hours.  We typically collect  400 data points per 3-4 second scan.  Prior to further analysis, we recast our frequency axis in terms of a normalized point number, $x_j  \equiv (j-N/2) / (N/2)$.  In this case $N=400$, and $1 \leq j \leq 400$, so that $-1 < x_j \leq +1$.

\begin{figure}[h]
\includegraphics[scale = 0.25]{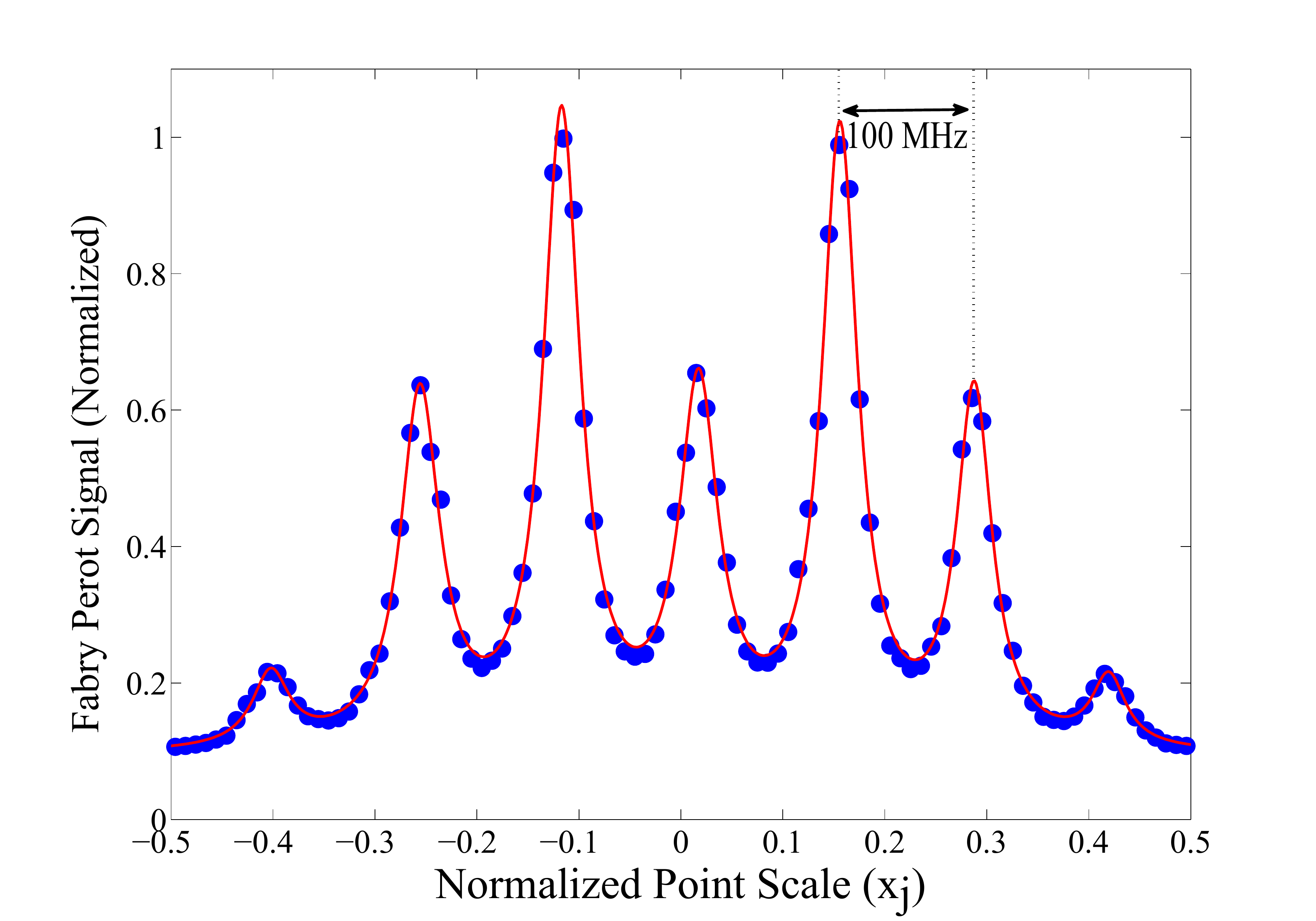}
\caption{A Fabry-Perot transmission data (points) with the results of a fit to a sum of seven Lorentzians (solid line).  The seven peak locations are exactly 100 MHz apart. All of these peaks correspond to a single FP longitudinal mode (free spectral range = 1.0 GHz).}
\label{fpscan}
\end{figure}

 We begin the analysis procedure by fitting the central FP spectrum.  Because we limit our laser scan to a single set of FM sidebands associated with a single FP longitudinal mode, we need not fit our FP spectrum to a periodic Airy function and instead fit the spectrum to a simple sum of seven Lorenztians.  We know that each Lorenztian peak is separated from the next by precisely 100 MHz, our modulation frequency.  However, due to nonlinearity in the PZT response, we expect that the fitted Lorenztian peak locations, $x^{(1)}\dots x^{(7)}$, will not have equal separation in terms of the variable $x$.  Given the relatively small frequency-scanning range, and the nonlinearity typical for such PZT devices, we find that a third-order polynomial allows an accurate frequency mapping of our scan.  Thus, to each point in our scan we assign a frequency according to $f(j) = a_0 + a_1 x_j + a_2 x_j^2 + a_3 x_j^3$, where the polynomial coefficients are determined by insisting that the Lorenztian FP peaks occur at exact 100 MHz intervals.  Fig. \ref{fpscan} shows the results of a fit to an FP spectrum. The deviation from linearity in our scan is small, but we find that quadratic and cubic components can be resolved.  These parameters tend to remain consistent for repeated scans taken under identical conditions.  Moreover, upward- and downward-going laser scans tend to have opposite signed quadratic components due to the hysteretic nature of the PZT.  We found that for these scans, a fourth order polynomial fit did not reliably resolve a quartic coefficient and did not improve the quality of the fit.   While this polynomial model may not be a reliable mapping of the frequency near the edges of the scan, where FP peaks do not exist, it does provide an accurate map of the frequency non-linearity in the central portion of the scan, exactly where the FM atomic beam signal is located.

In the $ \sim$14 seconds between HV on and off scans, a small change in the laser cavity PZT position or FP cavity length would obviously result in a systematic error in a Stark shift determination.  For this reason, we use the saturated absorption (SA) peak from our supplementary vapor cell as a frequency reference for both scans.   Using the newly calibrated frequency axis, we fit the central portion of the SA peak to a Lorenztian to determine the line center.  Independently, we can evaluate the extent of possible slow drifts in the FP or PZT systems by comparing the SA line center position to $a_0$ (derived from the FP fit) over time scales up to one hour.  We find that the maximum drift in relative line positions of the two signals to be 10 MHz over these time scales, and typically much less than this.  Nevertheless, as the Stark shifts we measure are of order 10-50 MHz, the use of the SA cell reference is clearly essential.  After the SA fit, we reset the frequency offset $a_0$ so that the SA peak frequency defines the frequency origin, while keeping the higher order  parameters $a_1 - a_3$ intact.  We can now compare the line center of our FM atomic beam spectra to the location of the SA peak using a properly linearized and calibrated frequency scale.

\begin{figure}[t]
\includegraphics[scale = 0.25]{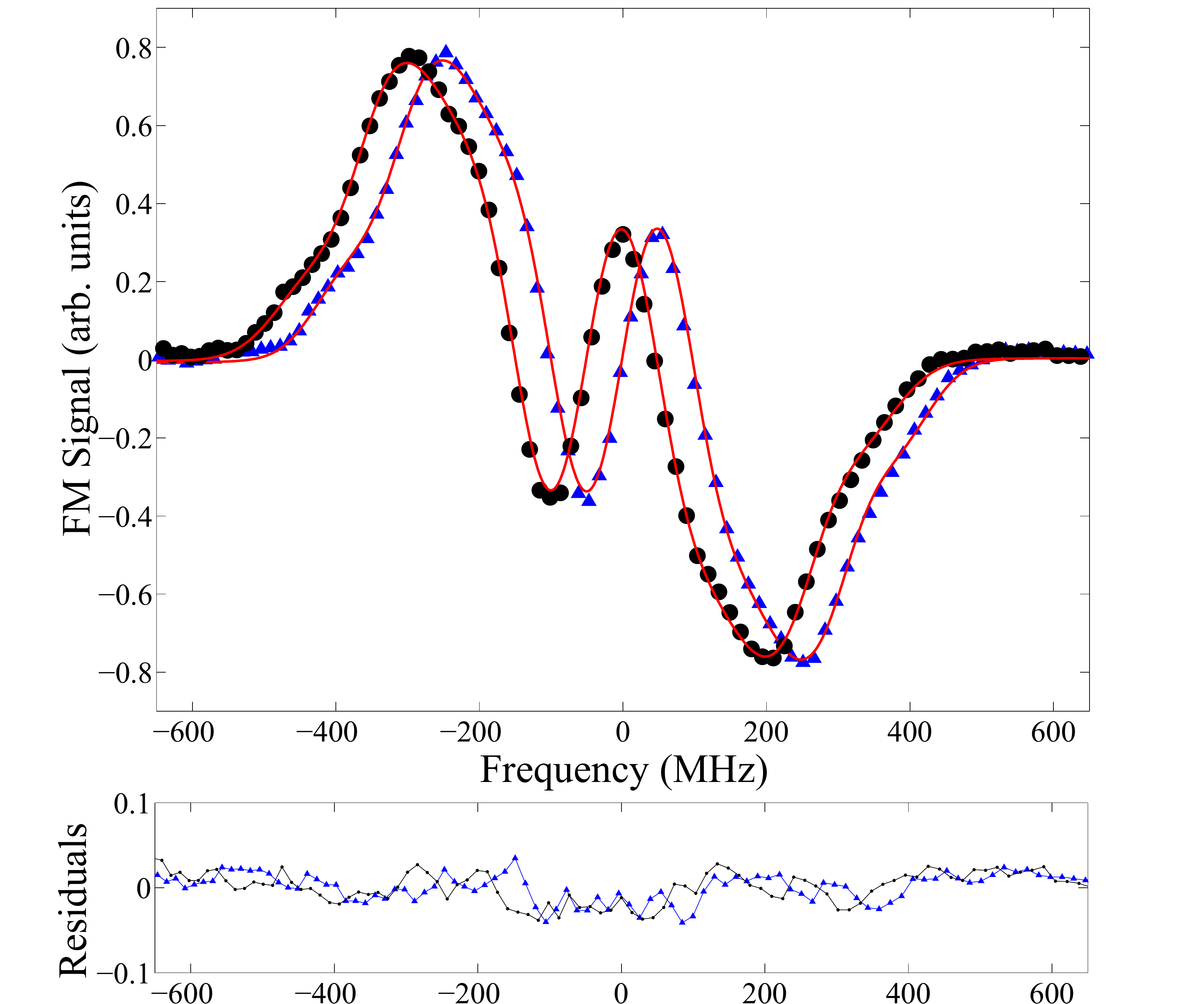}
\caption{Two up scan FM signals demodulated at 1f.  Triangles(circles) represent data taken with HV off(on).  Solid lines are fits based on the theoretical model discussed in the text.  Expanded residuals are shown below.}
\label{1fvoigtfit}
\end{figure}

\begin{figure}[t]
\includegraphics[scale = 0.25]{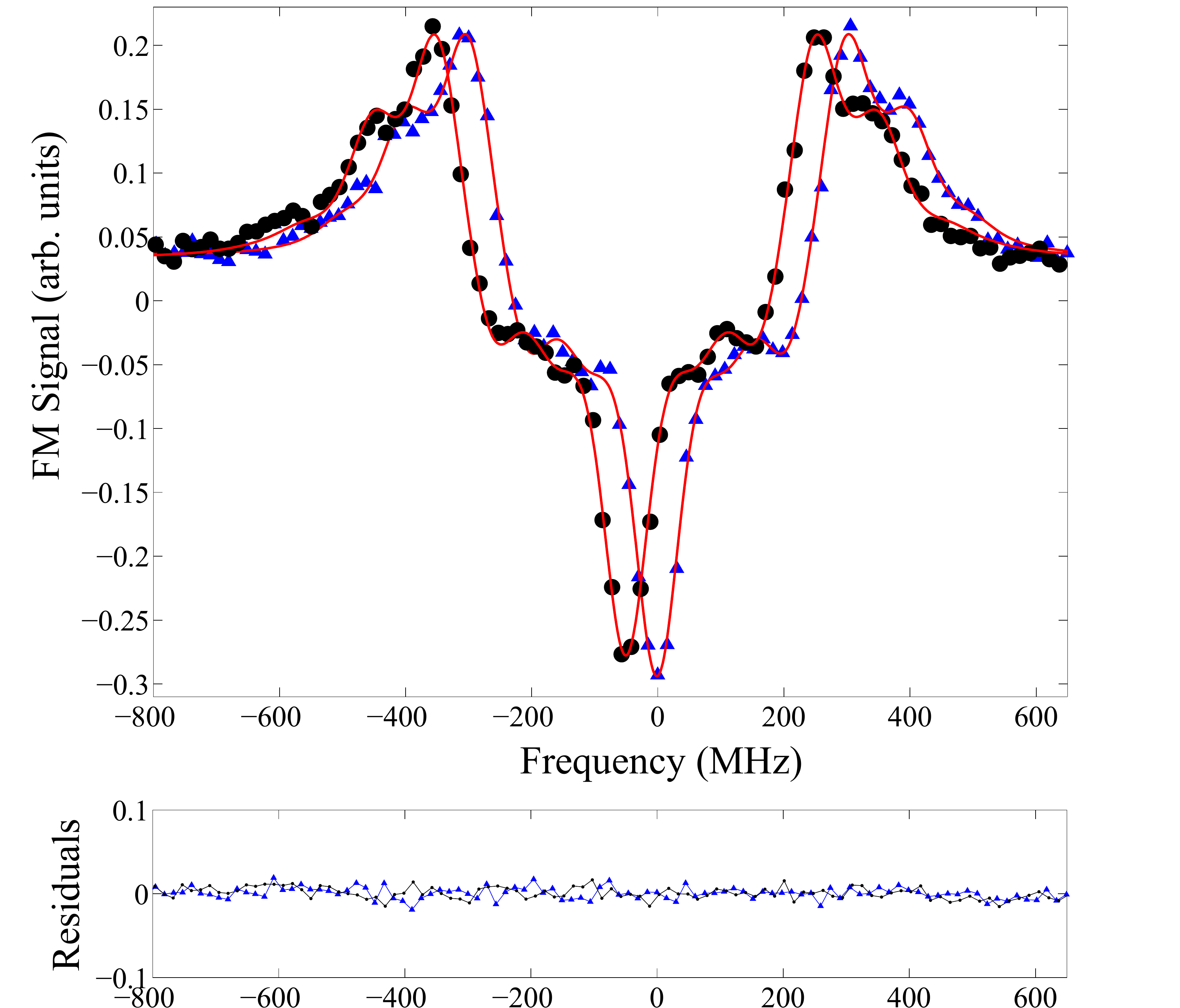}
\caption{Two up scan FM signals demodulated at 2f.  Triangles(circles) represent data taken with HV off(on).  Solid lines are fits based on the theoretical model discussed in the text.  Expanded residuals are shown below.}
\label{2fvoigtfit}
\end{figure}

The Stark shift is defined as the difference between the fitted center of the demodulated atomic beam transmission spectrum with the high voltage on and with the high voltage off (we observe a negative frequency shift upon application of the field).  If we demodulate the FM signal at 1f (2f) the resultant line shape is an odd (even) function.  At low modulation depths, the 1f signal has a simple dispersion shape, while the 2f signal consists of a symmetric peak with small additional lobes on each side of resonance.   For larger modulation depths, the demodulated signals grows in size, but also becomes substantially more complicated.  We have developed a curve-fitting algorithm based on functions derived from the full FM analysis outlined above.  That is, using Eq. \eqref{eq1f} as a basis for the fitting model, and truncating the infinite sum appropriately, we fit our demodulated data to a sum of symmetric and dispersive Voigt profiles, allowing the modulation depth, the relative phase of lock-in detection, and the Voigt profile component widths to become fit parameters in a non-linear least squares fitting routine.  Examples of these `full FM model' fits to our demodulated line shapes are shown in Figs. \ref{1fvoigtfit} and  \ref{2fvoigtfit}.  For both the 1f and 2f-demodulated cases, we show a pair of spectra with and without a 20 kV/cm electric field applied.  

While these full FM fits do an excellent job of capturing the complexities of the line shape, we also pursued a simpler, computationally faster fitting algorithm in order to assess the robustness of the line shape fitting results.  At  low modulation depths, we can fit the central portions of the 1f and 2f line shapes to  a linear function or a Lorenztian function respectively.  For somewhat larger values of $\beta$ one can develop a correspondingly more involved fitting function by fitting the central portion of the line shape to a higher-order polynomial, an odd function for the 1f signals and an even function for the 2f signals.  Since, as we observe, the pair of Stark-shifted line shapes are identical in size and shape, differing only in frequency offset, it is reasonable to assume that any imperfection in the polynomial approximation approach will affect each spectrum identically and thus \emph{not} lead to any systematic error in the Stark shift determination.  Furthermore, we can compare our results for 1f and 2f demodulation, and for low and high modulation depth to insure that Stark shift values we extract are not sensitive to the exact shape of the resonance line shape.   By exploring the degree of the polynomial fit, and the extent over which we fit the central portion of the line shape, we insured that the final Stark shift results were independent of these details.   The line shape centers and Stark shift constants extracted from the two different fitting methods were in excellent agreement.

\begin{figure}[h]
\includegraphics[scale = 0.5]{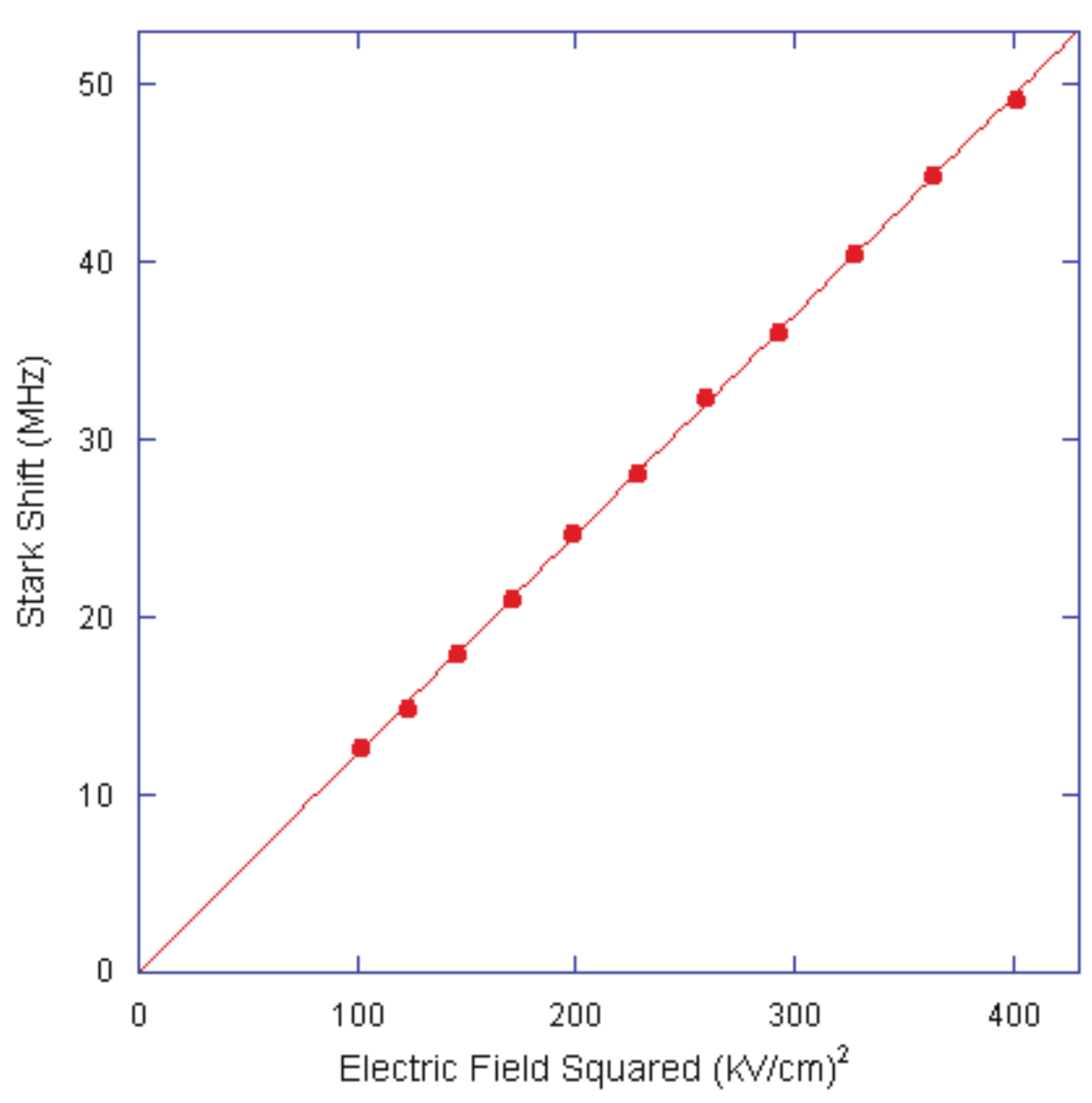}
\caption{Measured Stark shift vs. the square of the electric field. The result of one day's worth of data is shown, with all runs taken at each of 10 different nominal electric field values averaged. Statistical errors of individual points are comparable to the point size.    The Stark shift constant, $k_S$, (slope of fitted line) has an uncertainty of roughly 0.4\% for these data. }
\label{1DayData}
\end{figure}

\subsection{Analysis of Stark shift results}
\begin{figure}[h]
\includegraphics[scale = 0.65]{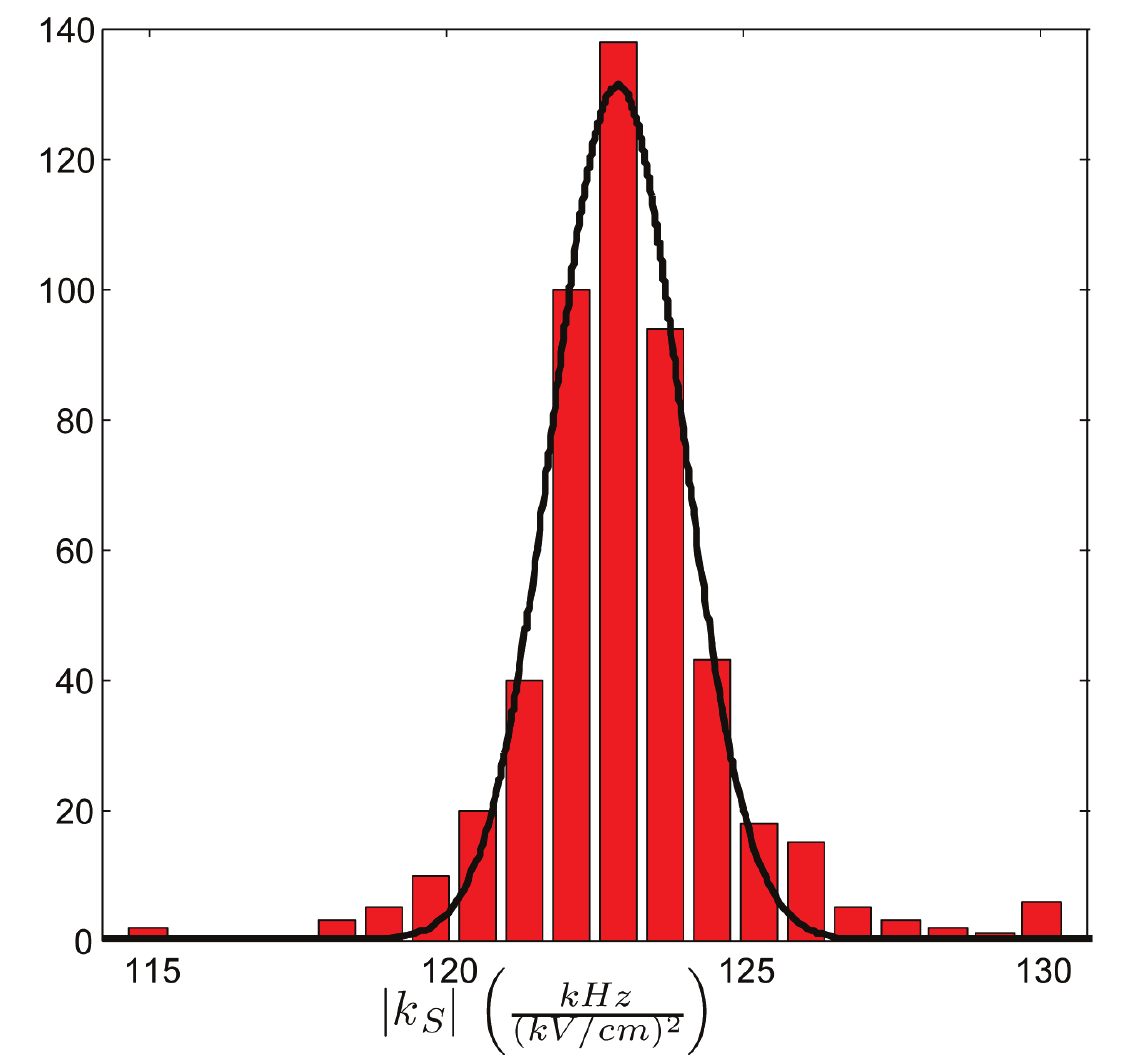}
\caption{A histogram of all $k_{S}$ values.  Each binned entry represents the results of a 30 minute collection cycle with nominally identical high voltage value. The solid line represents a fitted Gaussian.}
\label{TotalHistogram}
\end{figure}

A typical single Stark shift measurement, determined  from a pair of consecutive scans, yields a 3-5 MHz uncertainty, while the magnitude of the shift varies between 10 and 50 MHz for the HV values we use. We store results for upward laser sweeps (`up scans') and downward sweeps (`down scans') separately.  We further separate our Stark shift results into data sets based on the order of field switching -- that is, we have separate sets of on-off as well as off-on Stark shift results for each HV setting.  For each experimental condition, we compute a weighted average and standard error for the measured Stark shifts over one 30 minute cycle at a single HV value.  By dividing the resulting Stark shift value by the square of the electric field for this data set we can obtain a result for the Stark shift constant $k_{S} = \Delta\nu / E^2$, in units of kHz/(kV/cm)$^2$, which can eventually be converted to the atomic scalar polarizability.  Alternatively, after analyzing similar data sets for a full range of high voltages between 10 and 20 kV, we can plot our measured shifts vs. the square of the electric field, as is done in Fig. \ref{1DayData} for one complete day's worth of data.   It is encouraging to see the expected linear relationship, and that the linear fit yields an intercept value statistically consistent with zero.   When we extract a value for the Stark shift constant independently for each individual 30 minute cycle, for each value of high voltage, and under all experimental conditions, we obtain the histogram shown in Fig. \ref{TotalHistogram} consisting of roughly 500 distinct $k_S$ values.  We can obtain a final statistical mean and error from our data in several ways -- for example, by analyzing individual data such as in Fig. \ref{TotalHistogram}, or by computing slopes such as indicated in Fig. \ref{1DayData} for each day, and then averaging over all 18 days of data collection.  These various statistical analysis schemes gave consistent final results.  This then leads us to a final mean and one-standard-devation statistical error for our Stark shift measurement of:  $k_{S} = -122.92(05)$ kHz/(kV/cm)$^2$.

\subsection{Investigation of systematic errors}
\begin{figure}[h]
\includegraphics[scale = 0.55]{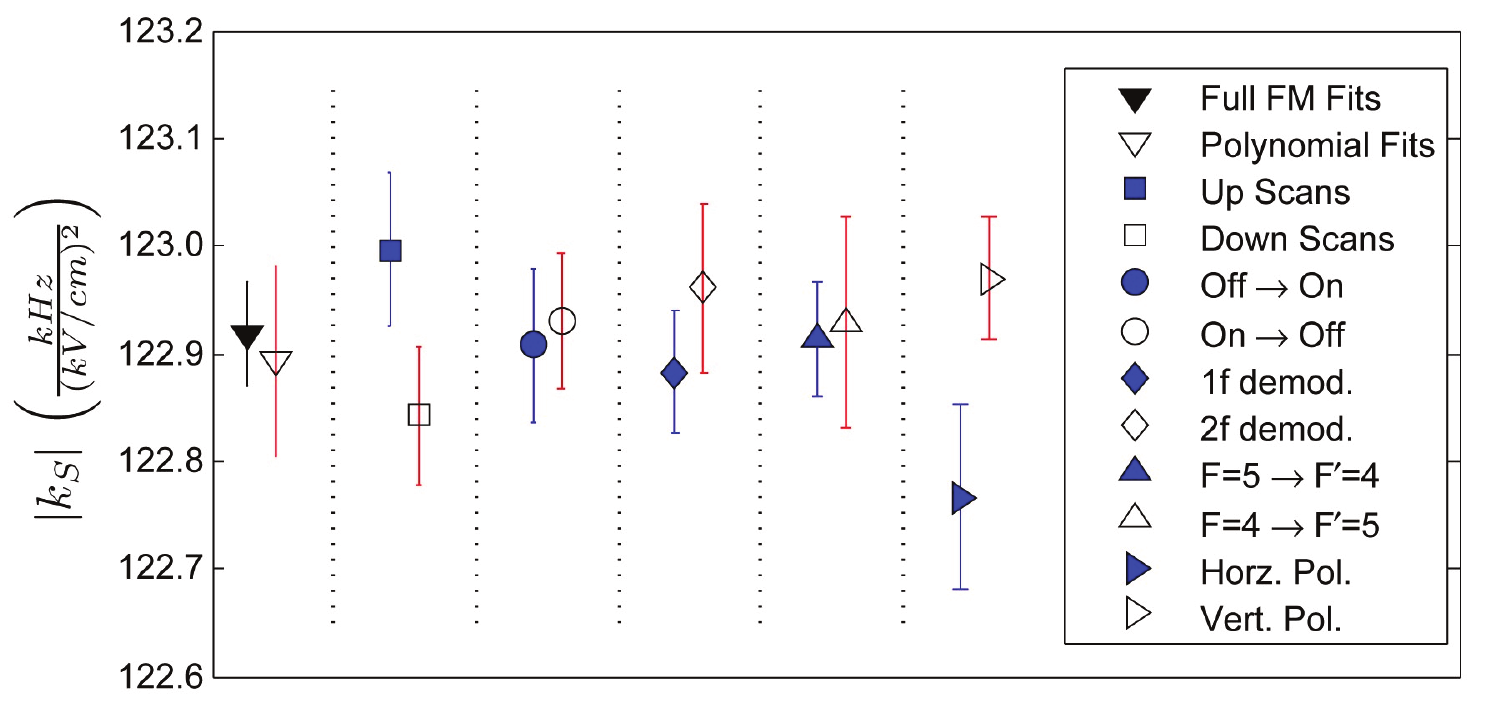}
\caption{A summary of data subset comparisons to investigate potential systematic errors.   The comparisons, in order, are:  fit results based on complete FM line shape analysis vs. those derived from polynomial or Lorenztian fits to the central portion of the spectrum;  direction of laser sweep;  sequence of HV switching; RF lock-in demodulation frequency;  hyperfine transition; and laser polarization.}
\label{2PComp}
\end{figure}

We searched for a wide variety of potential systematic errors in our data in two basic ways.  First, when two substantial subsets of the data could be identified, corresponding to two distinct experimental configurations, we compared the results of these subsets statistically.   Second, in some situations, we could search for correlations in our Stark shift constant values as a function of some experimental parameter.   In the first category, we compared data taken for up vs. down laser scans, for 1f vs. 2f demodulation, for $F= 4\rightarrow 5$ vs. $F=5\rightarrow 4$ hyperfine transitions, for horizontal vs. vertical laser polarization, and for HV on/off vs. off/on sequencing - none of which should influence our Stark shift constant. Fig. \ref{2PComp} summarizes these comparisons.      Most of these searches yielded no statistically resolved systematic differences.  Note that the entire vertical extent of Fig. \ref{2PComp} represents 0.5\% of our mean value.  For the cases of laser sweep direction and laser polarization,  where we see small ($\sim 2$ combined standard deviation) differences between subsets, we have included small systematic error contributions (see Table 1).

\begin{figure}[t]
\includegraphics[scale = 0.60]{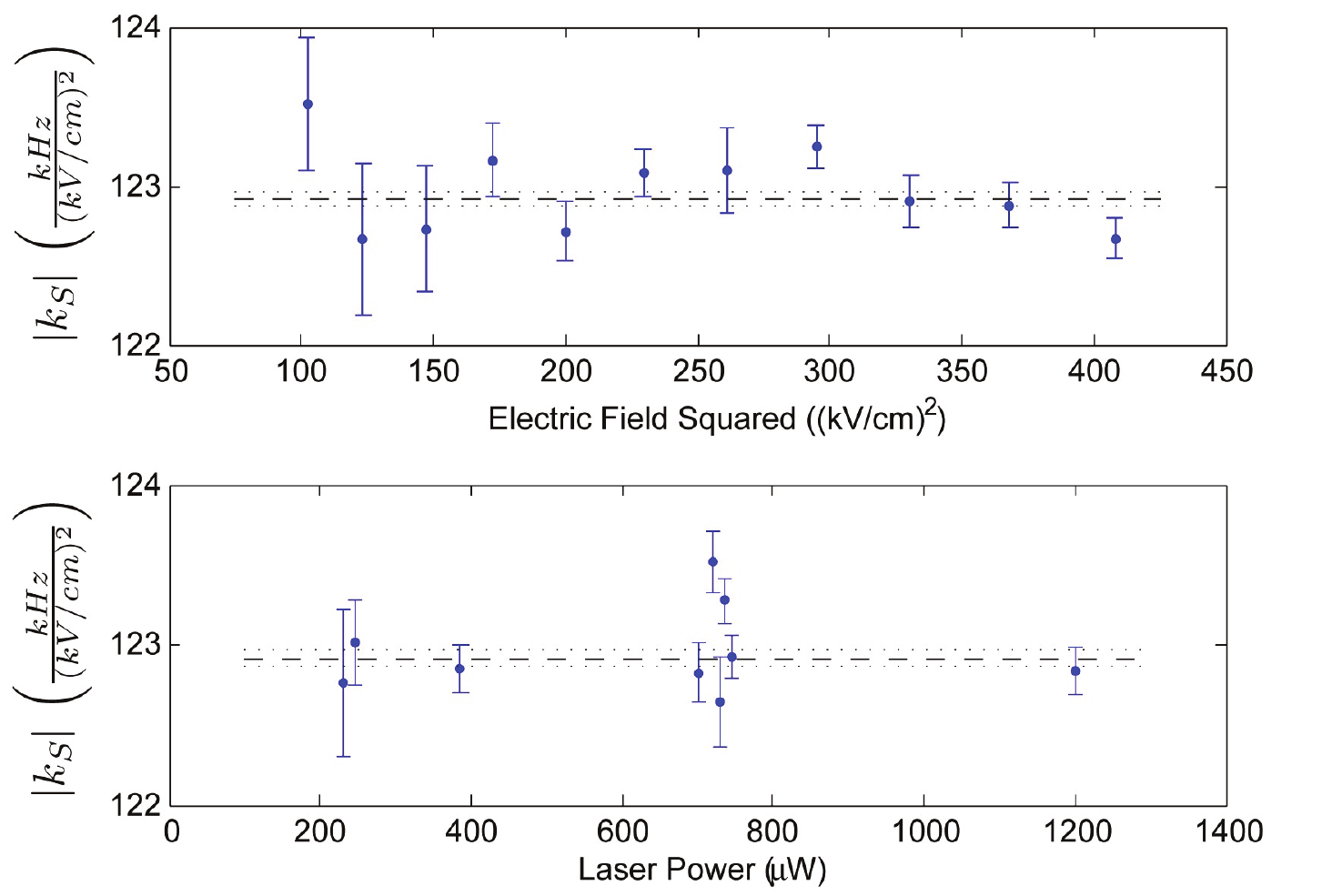}
\caption{Correlation plots:  $k_{S}$ versus the square of the electric field (top), and  $k_{S}$ versus laser power (below).}
\label{KvPower}
\end{figure}

We also searched for correlations in our data of the Stark shift constant with various experimental variables.  Two examples of this are shown in Fig. \ref{KvPower}.  At the top, we plot our measured values of $k_S$ versus  electric field (squared).  We see no statistically resolved slope here, but we do observe a scatter in values slightly larger than expected based purely on statistical variation (reduced $\chi^2 \sim 2$).  For this reason we have included a small additional systematic error contribution associated with the variation of $k_S$ with electric field in our final table of uncertainties.  At the bottom of Fig. \ref{KvPower} we plot $k_S$ vs. laser power.  One could worry that optical pumping or optical saturation effects or even AC Stark effects could affect our line shapes or line centers.  While it is difficult to see how this could affect the HV on/off scans differently causing a systematic Stark shift error, it is nevertheless encouraging to see that over a factor of six variation in laser power we see no observable correlation.   Finally, the list of error contributions in Table 1 includes a contribution from the overall uncertainty in the determination of the electric field (squared).  This is dominated by our 0.1\% uncertainty in the plate separation.  The other field-related uncertainties, from our high voltage divider and Keithley voltmeter, are an order of magnitude smaller.  We thus include a  0.2\% contribution to the final error from the uncertainty in $E^2$ determination.  As summarized in Table \ref{ErrorTable}, the quadrature sum of all errors, statistical and systematic, yields a 0.3\% final uncertainty.

\begin{table}[h]
\begin{centering}
\caption{Contributions to the final uncertainty.  All entries are in units of kHz/(kV/cm)$^2$.}
\renewcommand{\arraystretch}{1.5}
\begin{tabular}{|r|c|}
\hline
& \textbf{Sources of Error }\\ \hline \hline
 0.05&  \emph{Statistical} \\ \hline \hline
 0.05  &Electric Field Dependence \\ \hline
0.14 & Laser scan direction\\ \hline
 0.20  &Laser polarization \\ \hline
0.25  &E-Field Calibration \\ \hline \hline

\textbf{0.33} & \textbf{Quadrature Sum of Errors} \\ \hline \hline
\end{tabular}
\label{ErrorTable}
\end{centering} 
\end{table}

\section{Discussion of results and comparison with theory}
Our final result for the Stark shift constant is:  $k_{S} = -122.92(33)$ kHz/(kV/cm)$^2$.  Converting this result to scalar polarizability (in atomic units), we find that  $\Delta\alpha_0 \equiv \alpha_0( 6S_{1/2}) - \alpha_0(5P_{1/2}) = 1000.2 \pm 2.7$ $a_0^3$.  This result is in agreement with, but a factor of 30 more precise than, a measurement by Fowler and Yellin\cite{FowYel} in which they found $\Delta\alpha_0 = 944(73)$ $a_0^3$.  As discussed in \cite{Saf13}, a new theory calculation of this quantity has just been completed.  The result of this calculation gives $\Delta\alpha_0 = 995(20)$ $a_0^3$, in excellent agreement with our experimental result.  

 
The infinite sums in the theoretical calculations for the initial and final state polarizabilities are dominated (nearly 95\% of the total contribution) by terms involving mixing of the 6S$_{1/2}$ state and the nearby 6P$_{1/2, 3/2}$ states.  Because of this, we can combine the theoretical expression with our experimental result to infer highly accurate values for the matrix elements corresponding to these two transitions (and hence, the $6P$ excited state lifetimes).   Using summations as in Eq. \eqref{alphasum} as a starting point, we recast the expression for the polarizability difference to isolate the two dominant terms as follows:
\begin{eqnarray}
\Delta\alpha_0 & = & B \cdot S  + C,
\label{bsa}
\end{eqnarray}
where $S \equiv  \langle 6P_{1/2}\| D \| 6S\rangle^2$ (the E1 line strength), and we define
\begin{eqnarray}
B & = & \frac{1}{3} \Bigg( \frac{1}{E(6P_{1/2}) - E(6S)} + \frac{R^2}{E(6P_{3/2}) - E(6S)} \Bigg).
\label{bdef}
\end{eqnarray}
$R^2$ in Eq. \eqref{bdef} is the ratio of the $6S-6P_{3/2}$ to $6S-6P_{1/2}$ E1 line strength. In the present case, $R^2$ has the theoretical value\cite{Saf13} of 1.949(2), allowing us to compute a precise value for $B$.  The constant $C$  which accounts for all terms in the sums beyond the $6S-6P$ mixing terms, has the theoretical value $-60(8)$ $a_0^3$ in this case.  Inserting our experimental value into the left hand side of Eq. \eqref{bsa},  we can then infer a value for the line strength, $S$, with corresponding uncertainty (largely due to the theoretical error in $C$).  Finally, using the known energy splittings, we can compute a value for the decay rate of the two $6P$ states, using the relation\cite{SahooDasLifetime}
\begin{eqnarray}
A_{ab} & = & \frac{2.02613 \times 10^{18}}{\lambda^3}\frac{S_{ab}}{(2j_a + 1)} \textrm{ s$^{-1}$},\nonumber
\end{eqnarray}
where the line strength, $S_{ab}$, is expressed in atomic units, and the transition wavelength is in Angstroms.  In this case, the $6P$ excited state lifetimes are given by $\tau_a = 1/A_{ab}$, since there is only one E1 decay channel from each of these states.  We find
\begin{eqnarray}
\tau(6P_{3/2}) & = & 58.17(45) \textrm{ ns} \nonumber \\
\tau(6P_{1/2}) & = & 63.77(50) \textrm{ ns}. \nonumber
\end{eqnarray}
These values, with uncertainties below 1\%, are an order-of-magnitude more precise than earlier measurements derived from pulsed laser spectroscopy\cite{ZakiEwissSnoek6PLifetimes}, and agree very well with recent theoretical estimates\cite{SahooDasLifetime}.

\section{Conclusions and future work}
We have measured the scalar polarizability for the ground state 410 nm transition in atomic indium, improving the experimental precision by more than an order of magnitude.  This work has spurred a new set of atomic theory calculations whose results agrees very well with our measurement.  We are currently introducing a second (infrared) laser system to our experimental setup.  Using two-step excitation, we will then probe the $6S_{1/2} - 6P_{3/2}$ transition in our atomic beam with a diode laser centered at 1291 nm.  We recently completed an indium vapor cell measurement using this exact optical arrangement\cite{Gunawardena09}.  We plan to use frequency modulation spectroscopy again to probe the very weak IR absorption signal.  The Stark shift for this excited state transition is expected to be significantly larger, and will exhibit both scalar and tensor components, given the $J=3/2$ nature of the excited state. Since the $6P_{3/2}$ state polarizability is strongly affected by the presence of nearby D-states, this presents a new challenge to atomic theory. Finally, having developed this dual modulation scheme,  effective for studying weak transitions in the atomic beam environment, we plan to study directly the thallium $6P_{1/2} \rightarrow  6P_{3/2}$ M1 transition in the atomic beam.  Here, because the vapor pressure of thallium is much larger, a similar atomic beam source can produce a much denser downstream beam, which would help to compensate for the intrinsically small transition strength of the magnetic dipole line.

\begin{acknowledgments}
We thank Michael Taylor for important contributions in the design and development phase of this experiment. We wish to acknowledge the support of the National Science Foundation RUI program, Grant No. 0969781.
\end{acknowledgments}


\bibliography{Indium_bib}

\end{document}